\documentclass[prb,twocolumn,superscriptaddress,amsmath,amssymb,showpacs,floatfix,preprintnumbers,keywords]{revtex4-2}
\usepackage{amsmath}
\usepackage{amssymb}
\usepackage{graphicx}
\usepackage[export]{adjustbox}
\usepackage{epstopdf}
\usepackage{color}
\usepackage{epsfig}
\usepackage{braket}
\usepackage{amsmath}
\usepackage{stackengine}
\usepackage[colorlinks=true,citecolor=blue,linkcolor=blue]{hyperref}
\usepackage{amssymb}

\DeclareMathAlphabet{\bi}{OML}{cmm}{b}{it}

\begin{document}

\title{Coulomb impurity on a Dice lattice: atomic collapse and bound states}
\author{Jing Wang}
\email[]{wangjing@hdu.edu.cn}
\affiliation{School of Electronics and Information, Hangzhou Dianzi University, Hangzhou, Zhejiang Province 310038, China}
\affiliation{Departement Fysica, Universiteit Antwerpen, Groenenborgerlaan 171, B-2020 Antwerpen, Belgium}
\affiliation{NANOlab Center of Excellence, University of Antwerp, Belgium}

\author{R. Van Pottelberge}
\email[]{robbe.vanpottelberge@uantwerpen.be}
\affiliation{Departement Fysica, Universiteit Antwerpen, Groenenborgerlaan 171, B-2020 Antwerpen, Belgium}
\affiliation{NANOlab Center of Excellence, University of Antwerp, Belgium}

\author{Wen-Sheng Zhao}
\affiliation{School of Electronics and Information, Hangzhou Dianzi University, Hangzhou, Zhejiang Province 310038, China}

\author{Fran\c{c}ois M. Peeters}
\email[]{francois.peeters@uantwerpen.be}
\affiliation{Departement Fysica, Universiteit Antwerpen, Groenenborgerlaan 171, B-2020 Antwerpen, Belgium}
\affiliation{NANOlab Center of Excellence, University of Antwerp, Belgium}

\begin{abstract}
The modification of the quantum states in a Dice lattice due to a Coulomb impurity are investigated. The energy band structure of a pristine Dice lattice consists of a Dirac cone and a flat band at the Dirac point. We use the tight binding formalism and find that the flat band states transform into a set of discrete bound states whose electron density is localized on a ring around the impurity mainly on two of the three sublattices. The energy is proportional to the strength of the Coulomb impurity. Beyond a critical strength of the Coulomb potential atomic collapse states appear that have some similarity with those found in graphene with the difference that the flat band states contribute with an additional ring-like electron density that is spatially decoupled from the atomic collapse part. At large value of the strength of the Coulomb impurity the flat band bound states anti-cross with the atomic collapse states. 
\end{abstract}

\maketitle
\section{Introduction}\label{sec:1}
Since the experimental discovery of graphene as a stable two-dimensional allotrope of carbon the research into other 2D materials with different kind of band structures has experienced an increased interest~\cite{ref1}. This has led to many theoretical and experimental works on e.g. multilayer graphene, transition metal-dichalcogenides, gapped graphene, phosphorene, etc ~\cite{ref2,ref3}. Recently, the interest into materials with flat bands (dispersionless bands) increased with the discovery of the emergence of a flat band in twisted bilayer graphene ~\cite{ref4,ref5,ref6}. Since flat bands are (almost) dispersionless this results in a quenching of the kinetic energy allowing for almost perfect localized states ~\cite{ref7} and the emergence of different many-body states ranging from insulating to superconducting. 

One of the lattice structures of considerable interest is the $\tau _3$ or Dice lattice~\cite{ref8, ref9,ref10,ref11,ref12,ref13,ref14,ref15}. It can be viewed as the honeycomb lattice of graphene with an extra atom in the center. This lattice consists of two triangular sublattices where the hexagonal unit cell contains now an extra carbon atom that is coupled to one of the two sublattice sites (see Fig.~\ref{fig:fig1}). Because of this additional carbon atom the usual graphene Dirac cones are accompanied by a perfect flat band. Interestingly, the two body problem in graphene can be reduced to a particle living on a Dice lattice~\cite{ref16}. Initially the Dice model was mostly an interesting theoretical model, however recently materials have been discovered which can be modelled by such a Dice lattice~\cite{ref17,ref18,ref19}. Both the theoretical and experimental relevance of the Dice lattice remains to date a major motivation to study it in more detail.

Understanding how these flat bands are modified under the influence of external fields is of fundamental theoretical and experimental importance. For example, recently these flat band states were investigated in the case of nanoribbons with and without an applied perpendicular magnetic field ~\cite{ref20,ref21} showing their robustness to various kinds of fields or lattice modifications. Another interesting topic is the effect of a Coulomb charge on those flat bound states. The effect of a charged impurity on charge carriers in graphene has attracted a great deal of interest due to the prediction ~\cite{ref22,ref23} and detection ~\cite{ref24,ref25,ref26,ref27} of the atomic collapse phenomenon seen as the emergence of distinct resonances visible in the hole continuum. The effect of a charged impurity on the electronic states in a Dice lattice was recently investigated within the low energy continuum approximation as described by the Dirac-Weyl equation with pseudospin one ~\cite{ref28,ref29}. In such an approach, effects due to the discrete lattice are neglected which are preserved in a tight-binding approach. For a pseudospin-1 fermion system, Han et al.~\cite{ref28} obtained analytical solutions for the spinor wavefunctions and found a criterion for the occurrence of atomic collapse, $\beta>(L^2+1/4)^{1/2}$ that depend on the angular momentum $L$ ($\beta$ is the dimensionless strength of the Coulomb potential). In the case of graphene this condition is $\beta>L+1/2$. However, the effect of the Coulomb impurity on the flat band states was not investigated. Gorbar et al.~\cite{ref29} considered the gapped version of the Dice lattice. Within the continuum approach they found bound states that split from the flat band which ultimately turn into collapse states with increasing strength of the Coulomb potential. However, the occurrence of such discrete bound states in the case of the continuum model was found to be questionable and in Ref.~\cite{ref30} it was claimed that it should lead to a continuum of strongly localised states. Here, we will revisit this problem by properly taking into account the discrete lattice structure which turns out to be essential to correctly describe those flat band states in the presence of a Coulomb impurity. A numerical 'exact' calculation will be presented. We will show that this makes a fundamental difference in the effect of the flat band states on the atomic collapse states and in addition we find a discrete set of bound states resulting from the flat band.

 We found that for the Dice lattice a two fold story unfolds with increasing strength of the Coulomb potential: (i) sharp features in the local density of states (LDOS) emerge from $E = 0$ corresponding to flat band states which are spatially localized, (ii) these sharper resonances are accompanied by broader atomic collapse resonances at lower energy for sufficient large Coulomb potential strength $\beta$. Both phenomena should exhibit clear signatures in the energy resolved LDOS and spatial LDOS.  

\section{Model}\label{sec:2}
The Dice model consists of a graphene lattice with an additional carbon atom at the center of each hexagon. While the hopping parameters between the graphene lattice atoms itself remain unaltered the additional carbon atom at the center of each hexagon will be coupled to only one of the graphene sublattices. 

\begin{figure}
\includegraphics[scale=0.8]{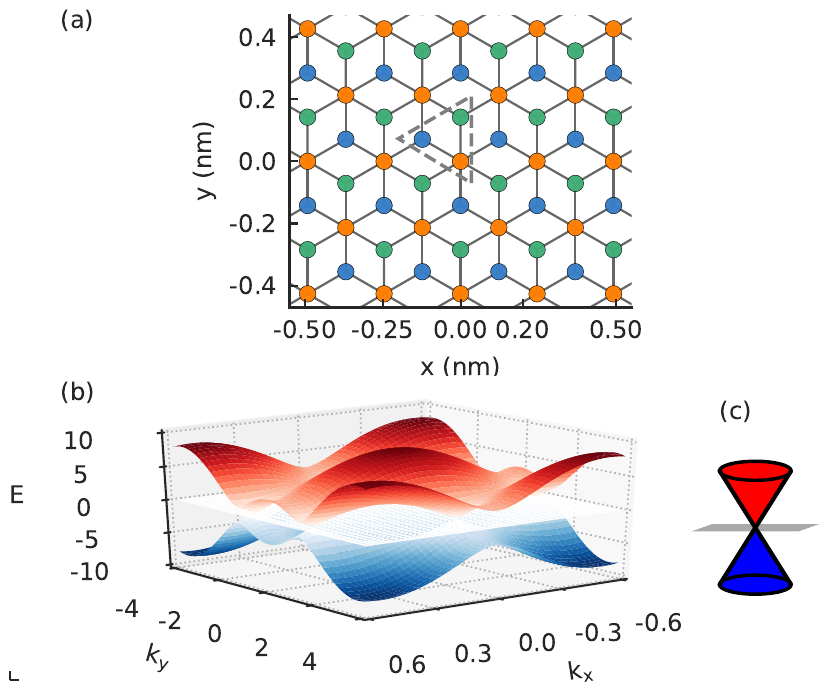}
\caption{\label{fig:fig1} (a)  Lattice structure of the Dice lattice where the blue and orange atoms refer to the graphene lattice and the additional carbon atoms at the center of each hexagon (green atom) is connected with only the blue atoms. In the Dice model all the hoppings between the atoms are taken the same. The dashed triangle shows the unit cell. (b) Energy spectrum of the Dice lattice. (c) Zoom into low energy dispersion around the K point.}
\end{figure}

In Fig.~\ref{fig:fig1}(a) we give a representation of the implemented lattice structure with $a=0.246$ nm the lattice constant. The graphene sublattice atoms (orange and green sites) are connected with each other via the usual graphene hopping parameter $t=-2.8$ eV. The additional carbon atom (blue sites) is connected with only one of the two sublattice atoms (here the orange sites) as indicated in Fig.~\ref{fig:fig1}, the hopping between this atom and the additional one is taken to be the same as the graphene hopping parameter. Due to this additional carbon atom the symmetry of the graphene lattice is dramatically altered: green and blue atoms have the same number of nearest neighbours (see Figs.~\ref{fig:fig1}(a)) while the orange atoms have twice the number of nearest neighbours. 

Bulk Dice lattice system is described by the following Hamiltonian
\begin{equation}
\hat{H}=\sum_{<ij>}t_{ij}c_i^\dagger c_j+\sum_{<il>}t_{il}c_i^\dagger c_l+h.c.
\end{equation}
The first term corresponds to the usual graphene tight-binding Hamiltonian and the sum $<ij>$ runs over all orange and green sublattice atoms shown in Fig.~\ref{fig:fig1}(a). The second term corresponds to the additional carbon atoms (blue sites in Fig.~\ref{fig:fig1}(a)) and the sum $<il>$ runs over all the blue and orange atomic sites indicated in Fig.~\ref{fig:fig1}. In general the couplings $t_{ij}$ and $t_{il}$ can be different but in the Dice lattice they are chosen equal. The $< >$ refers to the fact that only nearest neighbors interactions are include. The energy spectrum (see Fig.~\ref{fig:fig1}(b) and \ref{fig:fig1}(c)) consists of the typical graphene Dirac cone with in addition a flat band at the Dirac point.

We add an electrostatic potential to the free field Hamiltonian (1) which in our case is due to the presence of a Coulomb charge which we model by a Coulomb potential $V(r)=-\hbar v_F\beta/\sqrt{r^2+r^{*2}}$ with $\beta=Ze^2/\kappa \hbar v_F$ the dimensionless coupling constant, $Z$ the charge of the impurity, $\kappa$ the effective dielectric constant, $v_F$ the Fermi velocity, and we took $r^*=0.5\text{ nm}$ as a regularization parameter as done in Ref.~\cite{ref25}. The tight-binding Hamiltonian is solved using the open source tight-binding package Pybinding ~\cite{ref31}. This package solves the tight-binding Hamiltonian using the Kernel Polynomial Method (KPM-method). In all calculations a hexagonal flake with 200 nm sides is used. The size of the Dice lattice flake is taken large enough to ensure that locally bulk properties are achieved. We added an energy broadening of 10 meV. 

\begin{figure}
\includegraphics[scale=0.8]{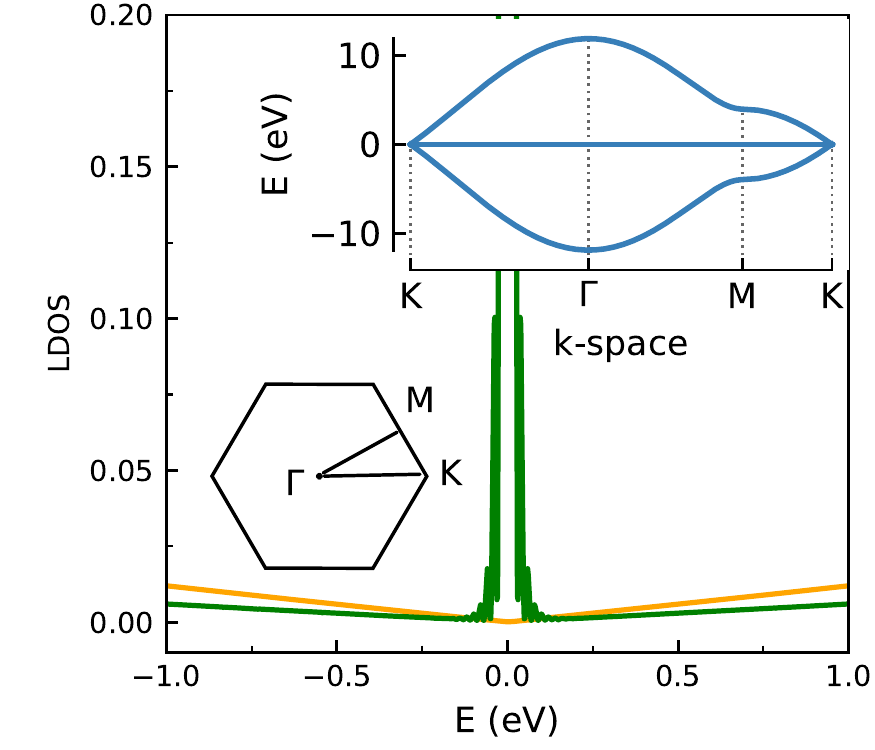}
\caption{\label{fig:fig2}LDOS for $\beta=0$ for atoms corresponding to the orange (orange curve) and green sublattices (green curve). The sharp peak corresponds to the flat band states and is only observed on the green and blue sublattices in Fig.~\ref{fig:fig1}(a). The upper insert is the band structure of the Dice lattice and the lower insert shows the first Brillouin zone.}
\end{figure}

\begin{figure}
\includegraphics[scale=0.8]{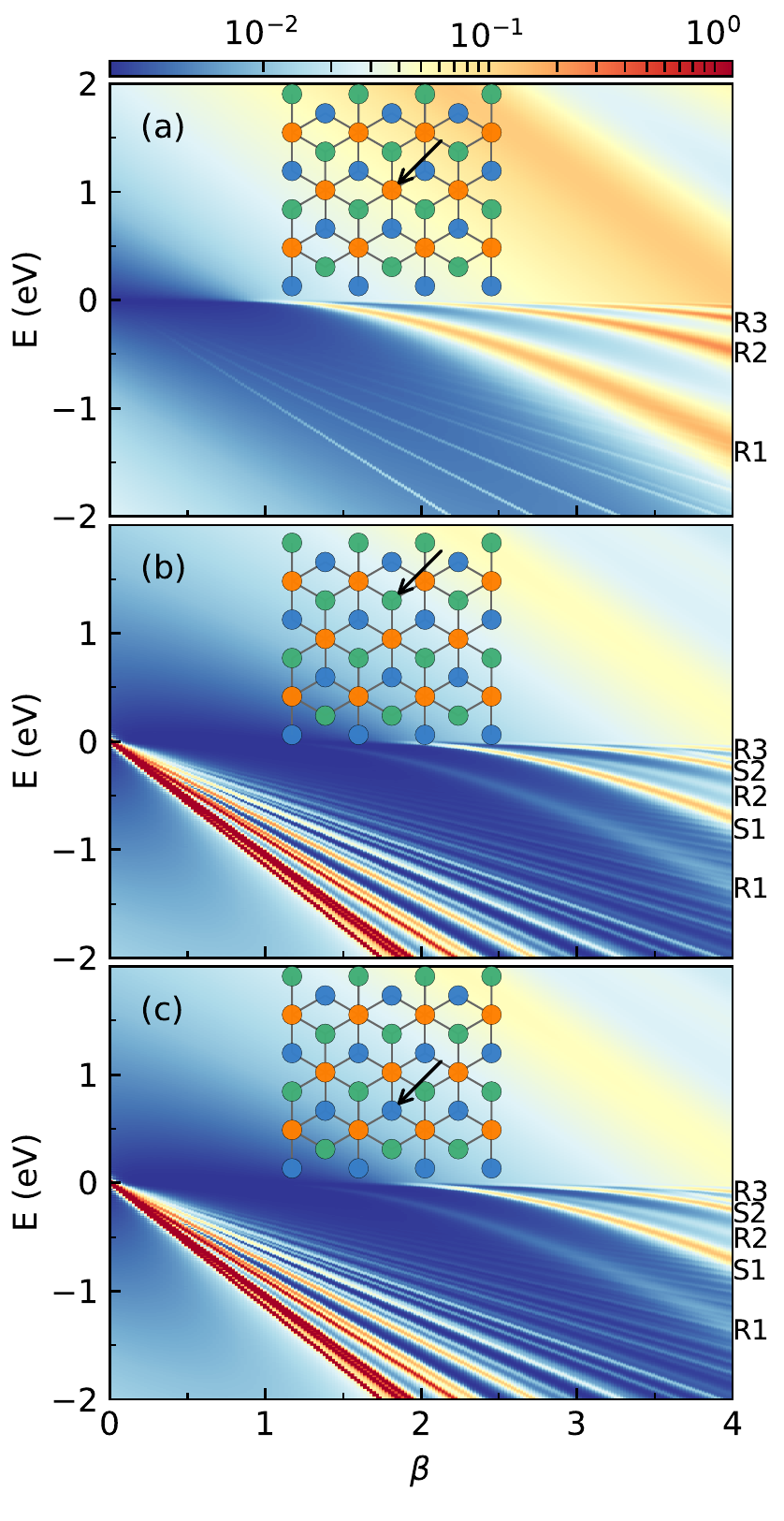}
\caption{\label{fig:fig3} LDOS calculated at three different positions of the Dice lattice (indicated by the black arrow in the inset of each figure). In (a) and (b) the LDOS is calculated at an atom corresponding to the graphene lattice (orange and green site) while in (c) the LDOS is calculate at the additional carbon atom (blue site).}
\end{figure}

\begin{figure}
\includegraphics[scale=0.8]{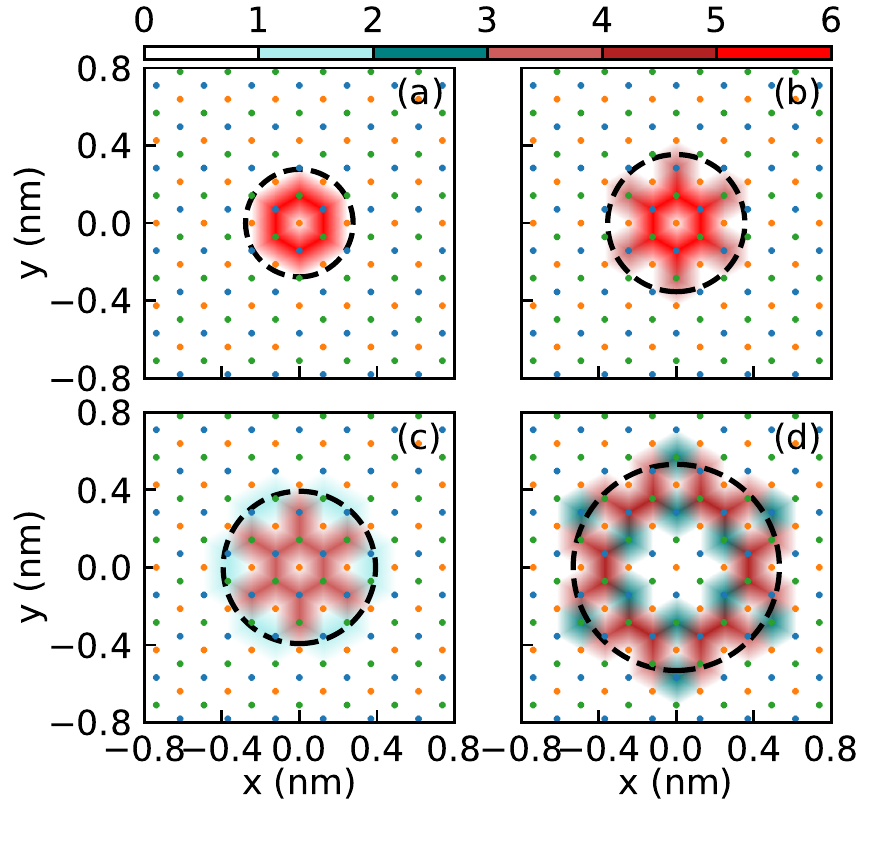}
\caption{\label{fig:fig4}Spatial LDOS for four lowest states originating from $E=0$ for $\beta=1.5$. The dashed circle indicates the radii of the Coulomb potential ($r_0$) at these energies. (a) $E=-1.73$ eV, $r_0=0.28$ nm, (b) $E=-1.61$ eV, $r_0=0.35$ nm, (c) $E=-1.55$ eV, $r_0=0.39$ nm, and (d) $E=-1.35$ eV, $r_0=0.53$ nm. The dots corespond to the Dice lattice where the same colour convention is used as in Fig. 1(a).}
\end{figure}

\begin{figure}
\includegraphics[scale=0.8]{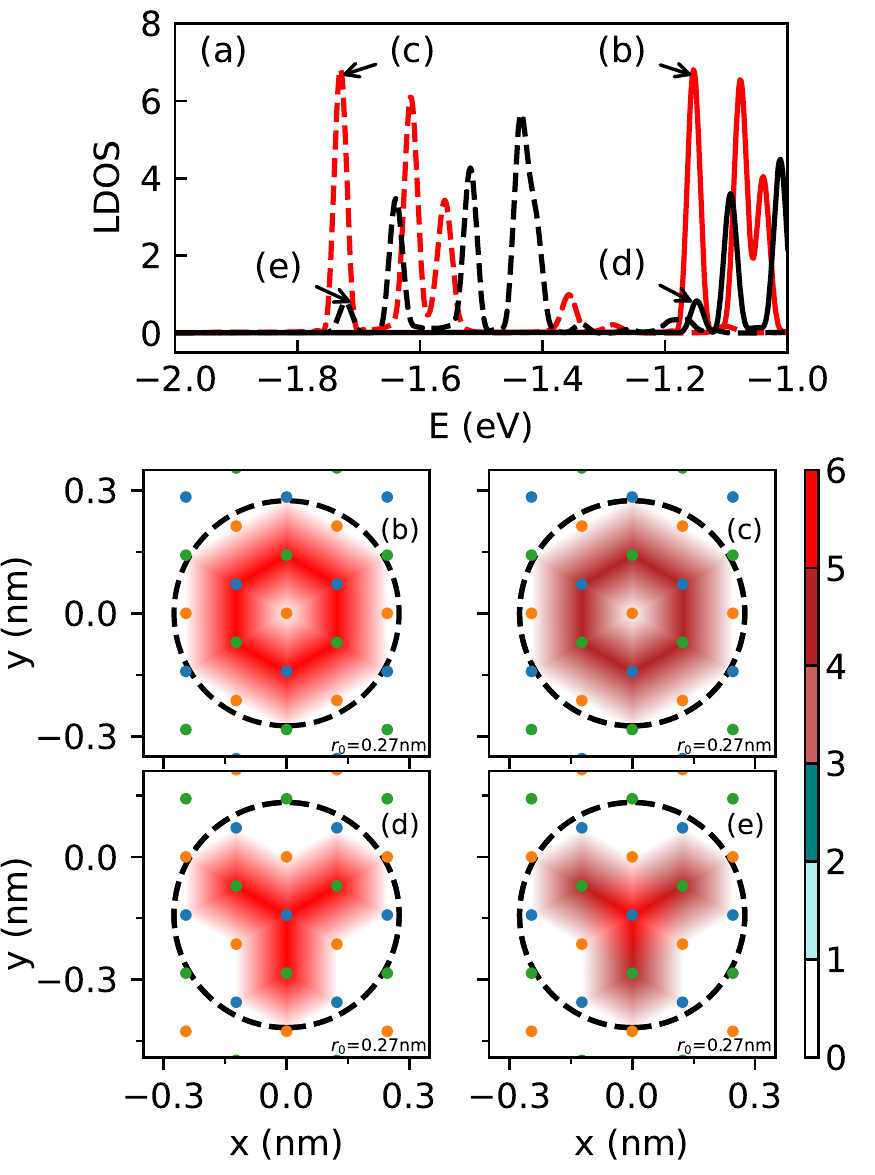}
\caption{\label{fig:fig5} (a) LDOS calculated for $\beta=1$ (solid line) and $\beta=1.5$ (dashed line). Red lines: Coulomb impurity is put on top of an orange atom and LDOS is calculated at a green atom. Black lines: Coulomb impurity is put on top of a blue atom and LDOS is calculated at a green atom. (b, c) The spatial LDOS is calculated for the same state for two different values of the charge indicated by the arrows in (a). (b) $E=-1.1535$ eV and (c) $E=-1.7294 $ eV. The dashed circle indicates the radius of the Coulomb potential at this energy. (d) and (e) show the same results as (b) and (c) but now the impurity is put on top of a blue atom and the hexagonal shaped density distribution changes in a triangular one.}
\end{figure}

\begin{figure}
\includegraphics[scale=0.8]{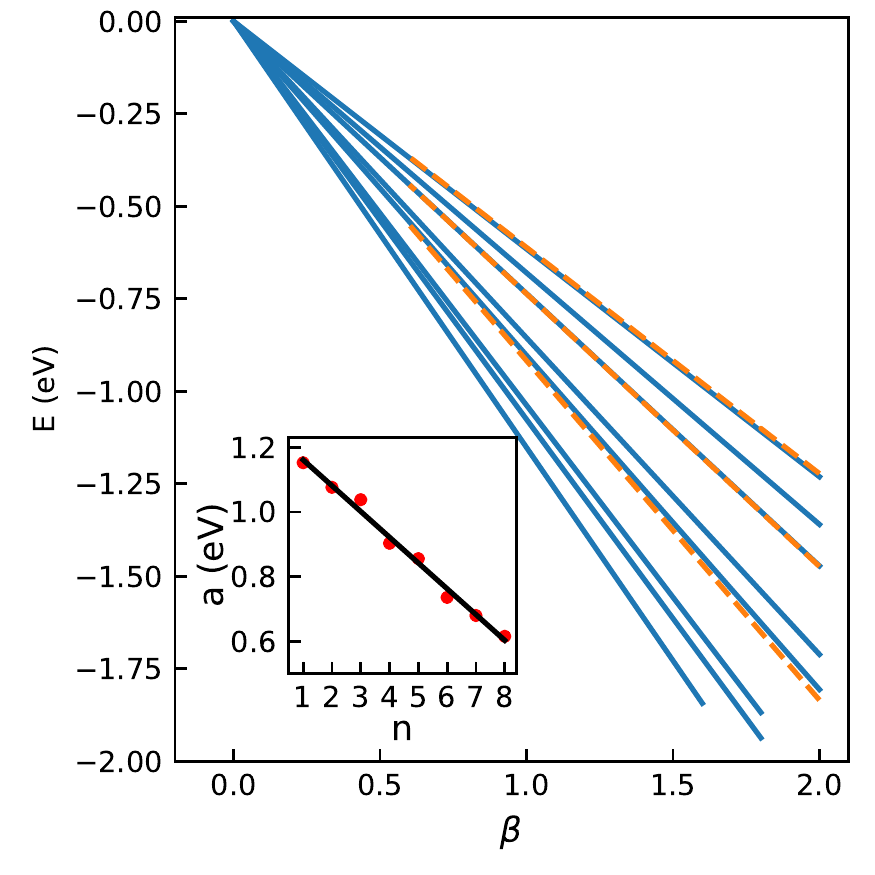}
\caption{\label{fig:fig6}First 8 lowest states originating from the flat band as function of the charge strength $\beta$. The orange dashed lines are from the LDOS taken at an orange lattice site. The blue solid lines are the LDOS at blue (green) lattice sites. The energy states behave as $E=-a\beta$. In the insert the value of the fitting parameter $a$ is shown for the different curves.}
\end{figure}

\section{Flat band bound states}\label{sec:3}
First we consider the pristine Dice ltattice (i. e. $\beta=0$) and calculate the LDOS at lattice sides belonging to different sublattices (see Fig.~\ref{fig:fig2}). For the LDOS at the green and blue atoms a peak in the LDOS for $E=0$ is observed corresponding to the flat band. For the orange sublattice atoms such a sharp peak at $E=0$ is absent and the LDOS shows only a linear behaviour as function of the energy like in the case of graphene. This result shows clearly that the flat band states are fully localized on the green and blue sublattices for $\beta=0$.

Next we calculate the LDOS at three different atomic sites as function of the impurity charge $\beta$ which is put at $(x,y)=(0,0)$ and thus on top of the orange atom. Fig.~\ref{fig:fig3}(a) shows the LDOS at atomic site where the Coulomb impurity is positioned and where flat band states are not localized for $\beta=0$. In Figs.~\ref{fig:fig3}(b) and \ref{fig:fig3}(c) we present the LDOS at atoms with flat band states for $\beta=0$. The first thing that can be noticed is that the LDOS in Figs.~\ref{fig:fig3}(b) and ~\ref{fig:fig3}(c) looks very different from the results shown in Fig.~\ref{fig:fig3}(a). The results in Figs.~\ref{fig:fig3}(b) and ~\ref{fig:fig3}(c) show a clear fan-like structure originating from $E=0$ which is visible for all values of the impurity charge $\beta$. This fan like structure is not visible for all values of the charge in Fig.~\ref{fig:fig3}(a) but becomes subtly visible for higher $\beta$ values. This behaviour can be understood from the fact that in Figs.~\ref{fig:fig3}(b) and ~\ref{fig:fig3}(c) the LDOS is calculated on a green and blue sublattice atom respectively, which for $\beta=0$ are identical and both lattice sites show a sharp peak in the LDOS located at $E=0$ that correspond to the flat band states as illustrated in Fig.~\ref{fig:fig3}. 

In order to understand the fan-like states originating from $E=0$ it is instructive to calculate the spatial LDOS for a few of such states. In Fig.~\ref{fig:fig4} we plot the spatial LDOS for the four lowest energy states (energy values given in the figure caption) originating from the flat band. Notice that these states are sharply localized in a hexagonal pattern around the impurity charge at the green and blue sublattices, reflecting a six-fold symmetry which originates from the lattice structure itself. The LDOS is localized within a distance from the impurity charge that equals the Coulomb potential radius at this energy (black dashed circle). The sharp localization itself can be explained by the nature of the flat band states. Since the flat band itself is dispersionless the kinetic energy is completely quenched and the Dirac equation is reduced to $V(r)\psi(r)=E\psi(r)$. This equation has as solution $\psi(x,y)\approx\delta(r-r_0)$ where the radius $r_0$ is determined by $V(r_0)=E$ as shown by the dashed circles in Fig.~\ref{fig:fig4}. Notice that this state is the 2D analogue of the 1D states recently found in a zigzag-nanoribbon~\cite{ref32} that was found to exhibit quasi-flat 1D bound states.

The fact that we observe a discretized spectrum originating from $E=0$ instead of a continuum of peaks as in the continuum model~\cite{ref30} can be explained from the fact that in the lattice model the localization is limited by the distance between the lattice size preventing a continuum of localized states that are theoretically possible in the continuum model ~\cite{ref28,ref29}. The linear $E\sim\beta$ dependence of these flat band states on $\beta$ can thus be understood from the sharp spatial localization of these states and the fact that the energy is given by $E=V(r_0)\sim\beta$. This results in only a very local effect of the Coulomb potential leading to a linear shift determined by the value of the Coulomb potential at that distance. In order to confirm this argument it is instructive to study the evolution of one of the flat band states with the size of the charge $\beta$. In Figs.~\ref{fig:fig5}(b) and ~\ref{fig:fig5}(c) we calculated the spatial LDOS for a particular flat band state, which is indicated by the arrows, for $\beta=1$ (red solid line) and $\beta=1.5$ (red dashed line) in Fig.~\ref{fig:fig5}(a). The qualitative behaviour of the spatial distribution of this state remains unchanged for the two different charges as shown by Figs.~\ref{fig:fig5}(b) and ~\ref{fig:fig5}(c), which is consistent with the linear $\beta\sim$ dependence of these states, i. e. $r_0$-value is only determined by the ratio $E/\beta$ which is the same in both cases. Notice that the relative amplitude of the three peaks for solid red and dashed red curves in Fig.~\ref{fig:fig5}(a) are the same. However, when the Coulomb impurity is put on top of a blue atom (solid (dashed) black curves for $\beta$ = 1 (1.5)) and we measure the LDOS also on a green atom, the number of peaks within the considered energy interval stays the same but their position and amplitude may differ. This is a consequence of the different coupling of the Coulomb potential with the different sublattic atoms which also results in a change of hexagonal to a triangular symmetry of the LDOS.

Fig.~\ref{fig:fig6} shows the energy of the fan originating from the flat band states as function of the impurity charge $\beta$, as derived from the LDOS in Fig.~\ref{fig:fig2}. The first 8 energy levels (blue lines in Fig.~\ref{fig:fig6}) with LDOS taken on the blue sublattice are fitted to $E=-a \beta$ with $a$ (eV) respectively 1.157, 1.086, 1.043, 0.904, 0.851, 0.719, 0.652 and 0.613. As discussed above, fewer flat band states are visualise for the LDOS calculated on the orange lattice which maybe a consequence of the much smaller amplitude of the LDOS. In the insert of Fig.~\ref{fig:fig6}, we show the fitting parameter $a$ as function of the number of the flat band state which shows on the average an almost linear dependence $a (eV)=-0.08n+1.239$. These states are localized on lattice sites (see Figs.~\ref{fig:fig4} and \ref{fig:fig5}) that are located at some distance from the Coulomb impurity. When drawing circles through lattice points it is clear that those circles are not equidistant which is the reason for deviations of the 'a'-parameter from a straight line shown in the inset of Fig.~\ref{fig:fig6}.

\begin{figure}
\includegraphics[scale=0.8]{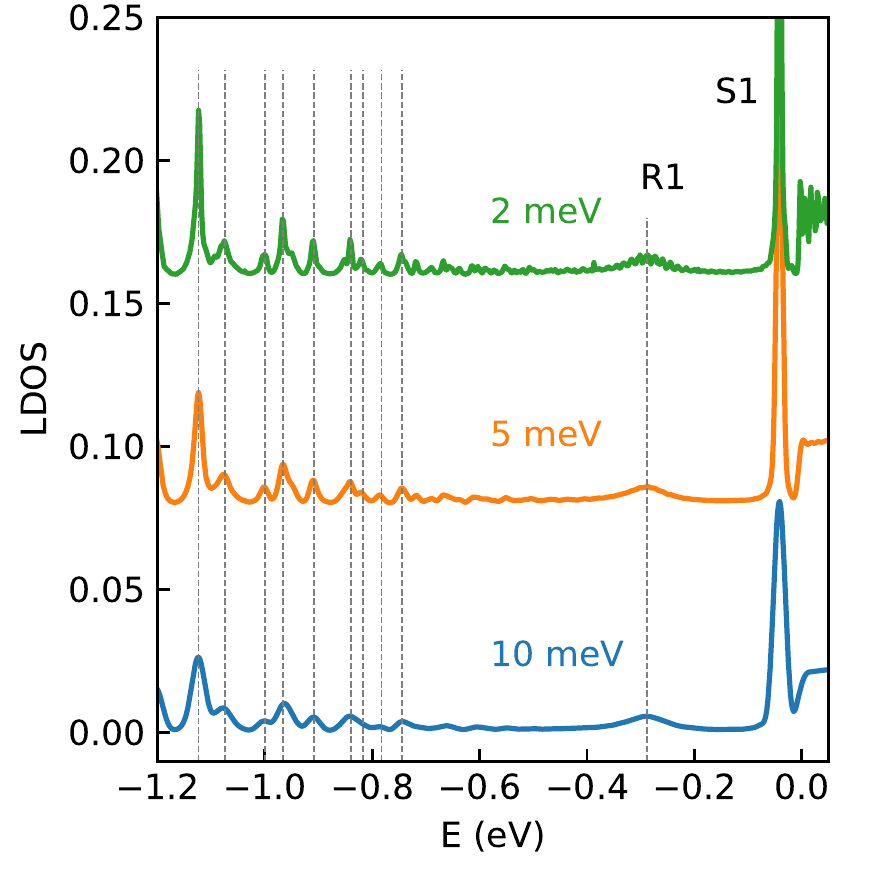}
\caption{\label{fig:fig7} LDOS calculated at the position $(x,y)=(0,0.142)$ for $\beta=2.2$ and for three values of the broadening parameter 10 meV, 5 meV and 2 meV. For this calculation a very large flake of  1400 $\times$ 1400 nm was used. Each result is offset with respect to the previous one for better distinction. With increasing broadening the width at 0.7 of the maximum of the left most flat band state is 0.0062 eV, 0.0126 eV and 0.0219 eV. The width of R1 state is 0.0624 eV, 0.068 eV and 0.07032 eV.}
\end{figure}

In all our calculations a fixed broadening of 10 meV was used. This broadening determines which features will be visible in the calculation, smaller broadening results in more details that can be resolved. However, larger flakes need to be simulated in order to prevent that finite size effect enters the results. Of course this broadening is also relevant in experiments where the energy level broadening is a relevant experimental parameter. It is therefore instructive to investigate how the broadening affects the flat band states. In Fig.~\ref{fig:fig7} we plot the LDOS for three values of the broadening parameter. With decreasing broadening the peaks in the LDOS that correspond to the flat band states increases and for smaller energies more flat band states can be resolved. These flat band states are bound states as their width increases linearly with the broadening parameter. However, the broader peak at $E=$ -0.3 eV corresponds to an atomic collapse resonance (R1 in Fig.~\ref{fig:fig2}) whose width is practically not affected by the broadening parameter as long as its intrinsic width is larger than the broadening. The width of a resonance state is determined by the life time of those states. 

\section{Atomic collapse states}\label{sec:4}

\begin{figure}
\includegraphics[scale=0.5]{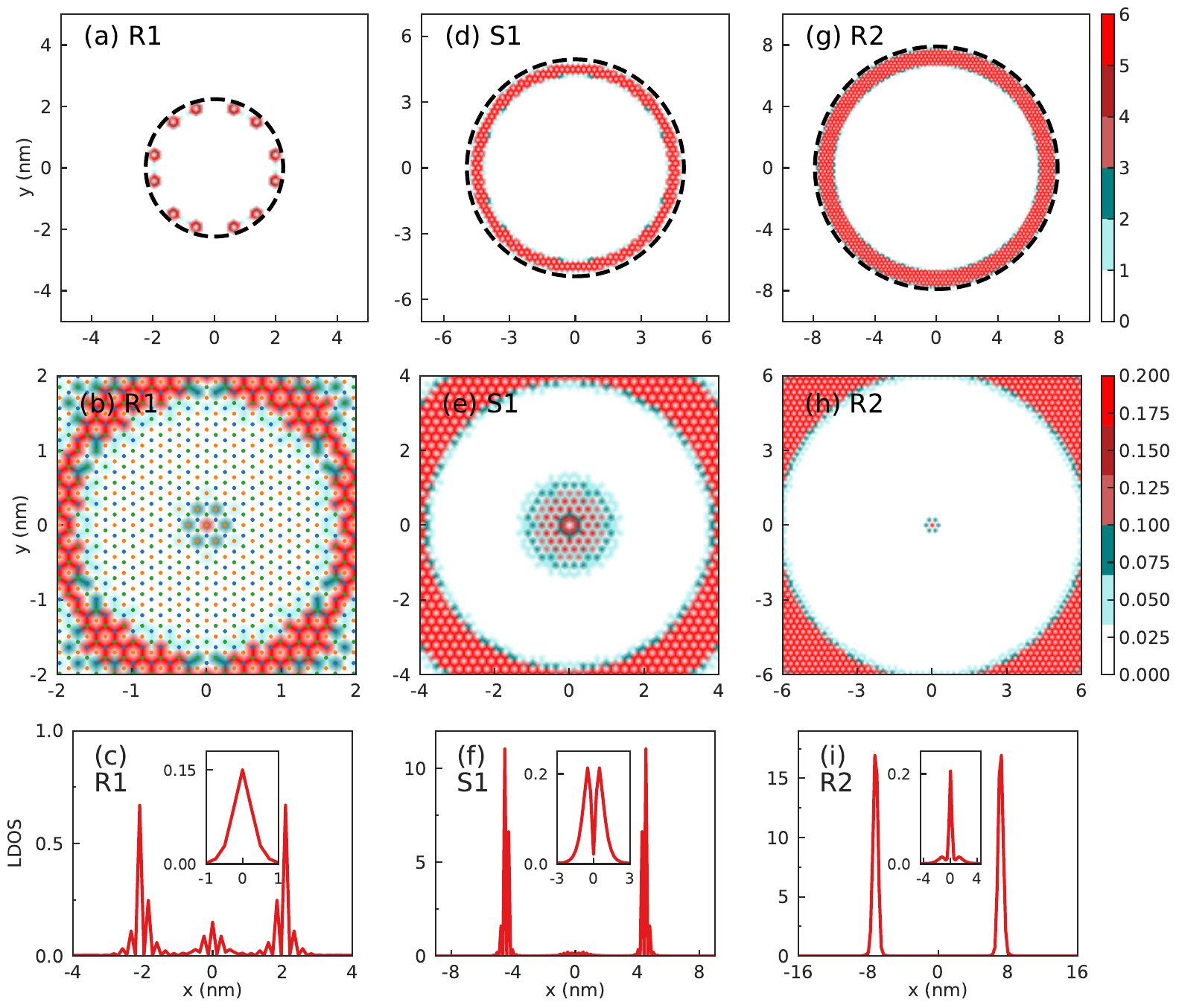}
\caption{\label{fig:fig8} Spatial LDOS for the resonances indicated in Fig.~\ref{fig:fig2} by R1, S1 and R2 for $\beta=3.5$. (a-c) $E=-1.00$ eV, $r_0=2.24$ nm, (d-f) $E=-0.46$ eV, $r_0=4.94$ nm and (g-i) $E=-0.29$ eV, $r_0=7.90$ nm. The dashed circle indicates the radius $r_0$ of the Coulomb potential at this energy. The second row are zooms of the top figures. The third row are cuts along $y=0$ where the inserts are enlargements around $x=0$.  }
\end{figure}

\begin{figure}
\includegraphics[scale=0.8]{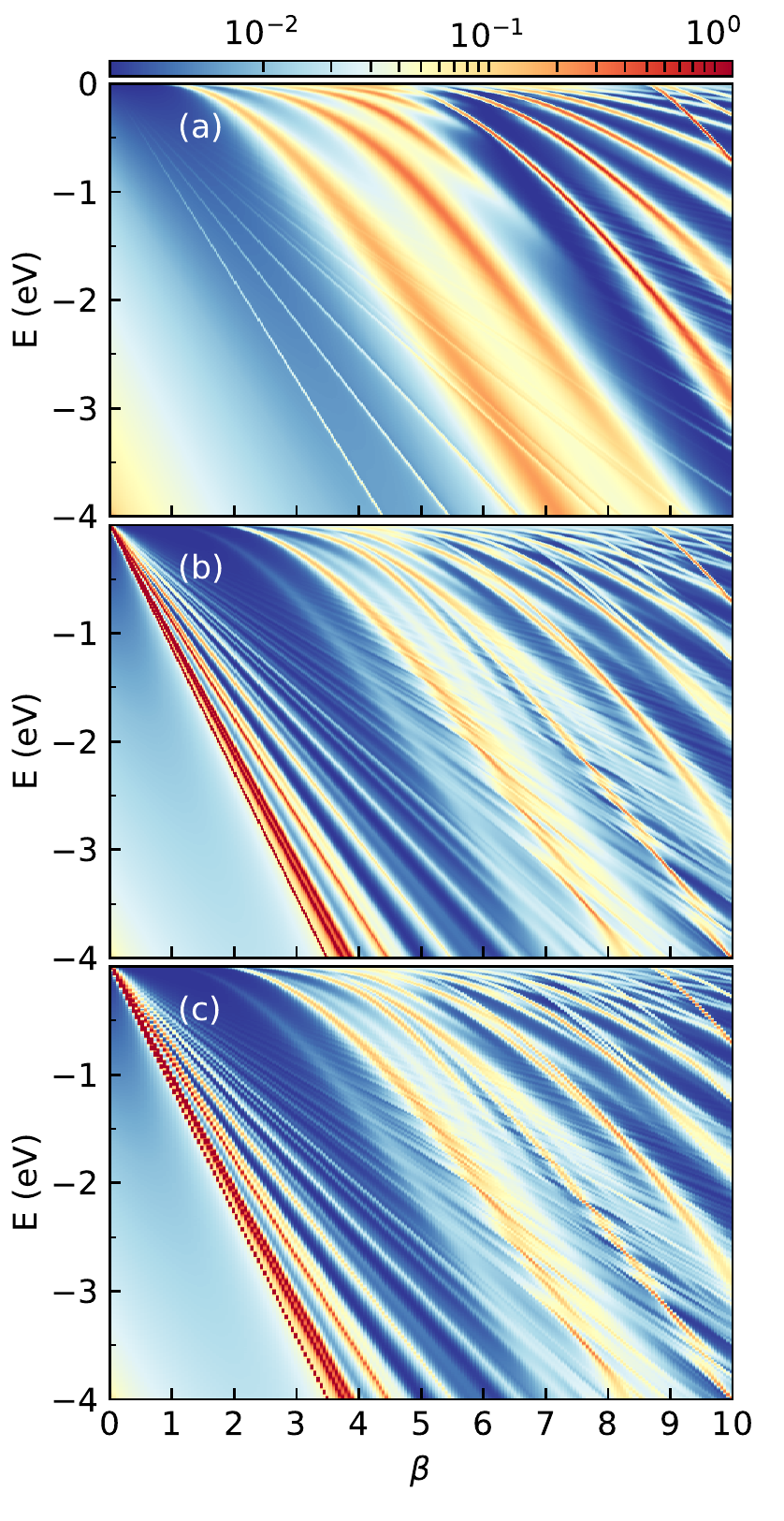}
\caption{\label{fig:fig9} The same as Fig. 2 but for a larger range of $\beta$ values.}
\end{figure}

\begin{figure}
\includegraphics[scale=0.8]{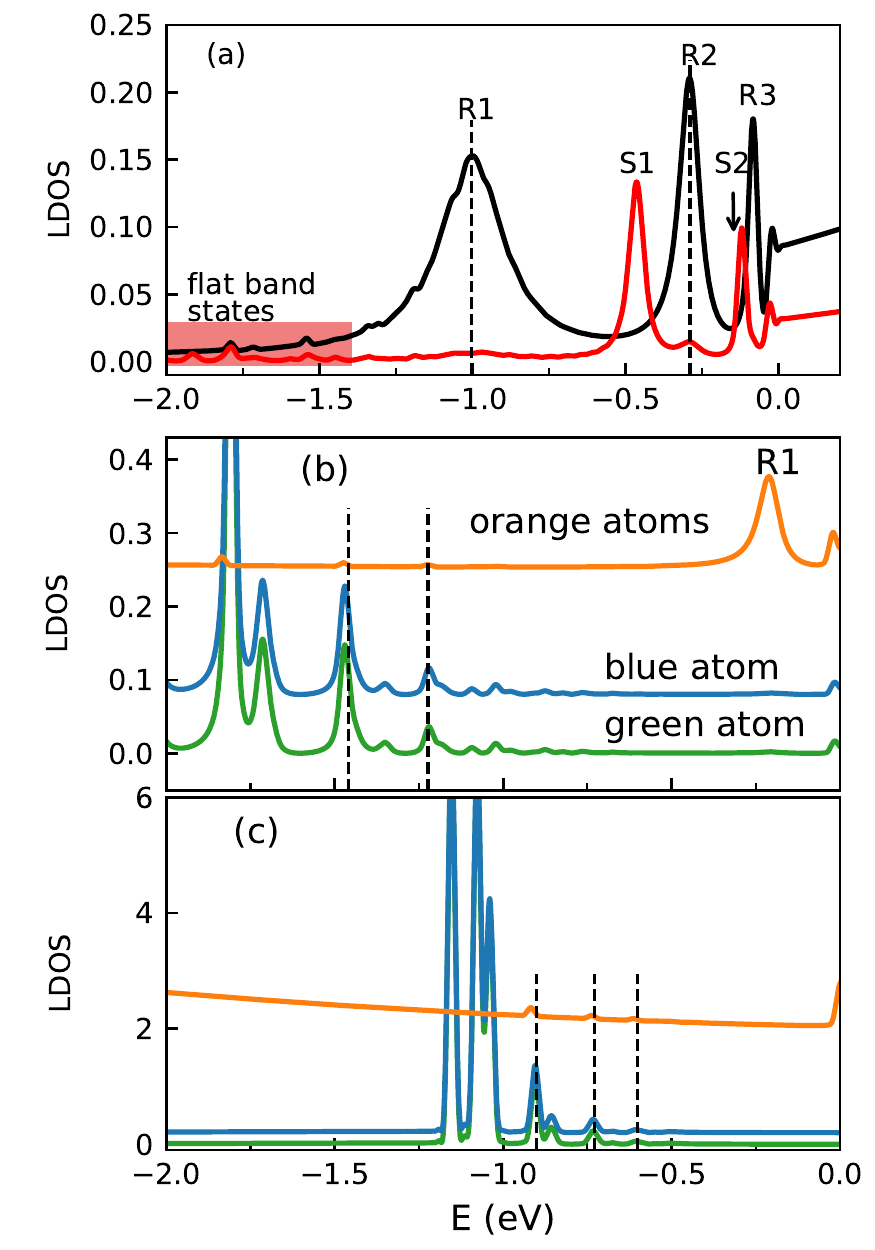}
\caption{\label{fig:fig10} Cut of the LDOS results shown in Fig.~\ref{fig:fig2} for (a) $\beta=3.5$. The black curve is a cut of Fig. 2(a)  and the red curve a cut of Fig. 2(b). Similar results are shown for (b) $\beta=2$ and (c) $\beta=1$.}
\end{figure}

Now lets focus on the broad resonant states (indicated by R1, S1 and R2 in Fig.~\ref{fig:fig2}) which show up for $\beta>0.5$. These resonances have a LDOS which are smaller than the states originating from the flat band and which are at least on order of magnitude more extended in space. These states correspond to the atomic collapse states. In order to confirm this and understand the nature of these states in more detail it is instructive to calculate the spatial LDOS for these states. In Fig.~\ref{fig:fig8} we plot the spatial LDOS for the three resonances indicated by R1, R2 and S1 in Fig.~\ref{fig:fig2} for $\beta=3.5$. The second row of the figure are enlargements of the first row. The lowest row are cuts of the spatial LDOS along $y=0$. The spatial LDOS (the first row of Fig.~\ref{fig:fig8}) exhibit a ring like structure. The ring corresponds to the contribution of the flat band states as discussed previously. But by zooming in on the spatial LDOS around the impurity, (1) for the R1 resonance (Fig.~\ref{fig:fig8}(b) and inset of Fig.~\ref{fig:fig8}(c)) a peak at the impurity position is found and the LDOS decreases along the radial direction, which is reminiscent of the $l=0, n=1$ hydrogenic orbital; (2) for the S1 resonance (Fig.~\ref{fig:fig8}(e) and insert of Fig.~\ref{fig:fig8}(f)) a node is found at the Coulomb center which is the counterpart of the $l=1, n=1$ hydrogenic orbital; and (3) for the R2 resonance (Fig.~\ref{fig:fig8}(h) and insert of Fig.~\ref{fig:fig8}(i)) a lower intensity ring outside the high LDOS center is found which therefore is a $l=0, n=2$ hydrogenic-like orbital.  

Note that in the case of the flat band states shown in Fig.~\ref{fig:fig4} no features corresponding to an atomic collapse state are observed at the center of the impurity. Thus in Fig.~\ref{fig:fig8} a spatial coupling between the atomic collapse states and flat band states is seen. However, the atomic collapse part of the state in the center is spatially seperated from the ring like electron density which occurs at much larger distance from the impurity. Notice that, in Ref.~\cite{ref28} the S-like resonances was not studied. Because we use the tight-binding Hamilton and do not have any index related with the angular momentum ($L$) as in the continuum approach of Ref.~\cite{ref28}, the crossing of atomic collapse with $L=1$ and $L=0$ is not show in Fig.~\ref{fig:fig2}. However, we found that for sufficient larger values of $\beta$ there is a anti-crossing between the flat bound states and the atomic collapse states as illustrated in Fig.~\ref{fig:fig9}.

With regards to future experiments it is also instructive to present a cut of the LDOS results shown in Fig.~\ref{fig:fig2} for a particular values of $\beta$. In Fig.~\ref{fig:fig10} we present a cut of the LDOS shown in Fig.~\ref{fig:fig2} for $\beta=3.5$, 2 and 1. The three atomic collapse resonances are labelled and form a broad clear signature in the LDOS in Fig.~\ref{fig:fig10}(a). The signature of the flat band states is also clearly visible as a number of distinct and sharp peaks showing up at larger negative energies. Both the flat band states and atomic collapse states present clear and distinct signatures visible in the LDOS. For those flat band states as shown in Figs.~\ref{fig:fig10}(b) and \ref{fig:fig10}(c), the LDOS calculated at blue and green sublattice atom have the same number of peaks while it has fewer and much weaker peaks at the orange sublattice atom. That's because the number of nearest neighbor interactions of the orange sublattice is different from the blue and green sublattice as shown in Fig.~\ref{fig:fig1}(a). All the flat band states observed in orange sublattice can be found in the LDOS calculated at the blue and green sublattice and are found at the same energy.   

\section{Conclusion}\label{sec:5} 
We showed how atomic collapse manifests itself in a Dice lattice. We found that the atomic collapse story in a Dice lattice is much richer than in graphene. The presence of a flat band in the Dice lattice results in strongly localised bound Coulomb states that have a spatial ring shaped LDOS. The energy of these bound states scales linearly with the strength of the Coulomb potential which is different from the traditional hydrogen problem where a quadratic scaling is found. The atomic collapse resonances have some similarities with those in graphene but are accompanied by ring shaped LDOS features that originate from the flat band states. We showed that both phenomena should pose clear signatures in both the LDOS and spatial LDOS in future experiments. Our results will be of relevance for different systems exhibiting a (nearly) flat band and for materials that can be described by a Dice lattice Hamiltonian. 

Lastly we would also like to note that here we applied a regularized Coulomb potential. It is however known from previous publications that the physics is not sensitive to the exact form of the Coulomb potential (only quantitatively) ~\cite{ref24,ref25,ref26,ref27}. Similar physics can be expected in the case of charged vacancies ~\cite{ref25} or in the case of potential wells created by an STM-tip ~\cite{ref26}. 

\begin{acknowledgments}
We thank Matthias Van der Donck for fruitful discussions. This work was supported by the National Natural Science Foundation of China (Grant Nos. 62004053, 61874038 and 61704040), the scholarship from China Scholarship Council (CSC: 201908330548) and the Research Foundation of Flanders (FWO-Vl) through an aspirant research grant for RVP.
\end{acknowledgments} 

\bibliography{references}
\end{document}